\documentclass[11pt]{article} 
\usepackage{amsmath,amsfonts,graphicx,mathrsfs} 

\newcommand{\be}{\begin{align}}
\newcommand{\ee}{\end{align}}
\newcommand{\bear}{\begin{eqnarray}}
\newcommand{\eear}{\end{eqnarray}}

\newcommand{\tr}{\mathrm{Tr}}
\newcommand{\sdet}{\mathrm {SDet}}

\newcommand{\ba}{\begin{array}}
\newcommand{\ea}{\end{array}}

\newcommand{\nn}{\nonumber}
\newcommand{\diag}{\textrm{diag}}

\begin{document}

\title{O$(N)$ colour-flavour transformations and characteristic polynomials of real random matrices}
\author{Yi Wei$^1$ and Boris A Khoruzhenko$^{1,2}$ \\  \\
$^1$ Isaac Newton Institute for Mathematical Sciences, Cambridge,UK\\
$^2$ Queen Mary University of London, School of Mathematical
Sciences, London, UK}
\date{}
\maketitle \abstract{The fermionic, bosonic and supersymmetric
variants of the colour-flavour transformation are derived for the
orthogonal group. These transformations are then used to calculate the
ensemble averages of characteristic polynomials of real random
matrices.}

\section{Statement of main results}
Since the pioneering work of Wigner, many physical systems have been
successfully studied with the help of random matrix models. Among
these asymmetric real random matrices arising in applications in
neural networks \cite{sommers}, quantum chaos \cite{efe} and QCD
\cite{jac} are known to be the most difficult.  In fact, until the
very recent breakthrough \cite{4,5,6,7,8}, the eigenvalue
correlation functions of real and complex eigenvalues were not
accessible even for Gaussian matrices and calculating the ensemble
averages of eigenvalue statistics in the complex plane for a
sufficiently general class of real random matrices remains a
challenging problem. The mathematical difficulties in calculating
the eigenvalue correlation functions of real random matrices are
mainly due to the fact that their eigenvalues are either real or
pairwise complex conjugate and the mathematical tools available to
study real random matrices are very limited, especially when
compared to those available for complex matrices. In this paper we
derive several integral transformations dealing with integrations
over real orthogonal matrices which we believe might be useful in
the above context. These integral transformations are known under
the name of the Colour-Flavour Transformations.

The Colour-Flavor Transformations (CFT) are certain types of
integral transformations based on Howe's dual pair theory. They were
was first derived by Zirnbauer \cite{cft} in 1996 and since then
have became a standard tool in mesoscopic physics, random matrix
theory, lattice QCD as well as other fields. The CFTs were
originally derived for U$(N)$ and Sp$(2N)$, and later generalized to
other classical groups \cite{bt,my,nagao,zirn}. However, both the
fermionic and bosonic variants of the O$(N)$ CFT appearing in the
literature \cite{nagao, zirn} do not seem to be reproduced in the
correct form, possibly suffering from typos. In this paper, we first
correct the bosonic and fermionic versions of the O$(N)$ CFT
and then derive the supersymmetric version, which is a new result. \\

\bigskip

\noindent{{Fermionic O$(N)$ CFT}}
\begin{align}\label{eq:fcft}
\int_{\mathrm{O}(N)}dO\exp(\bar{\psi}^i_a O_{ij} \psi^j_a)=C_0^F\int_{Z=-Z^{\mathrm{T}}}
d\mu(Z,Z^\dagger)
\exp\frac{1}{2}\bigg(\bar{\psi}^i_aZ_{ab}\bar{\psi}^i_b+\psi^i_aZ_{ab}^\dagger\psi^i_b\bigg)\;,
\end{align}
where $C_0^F$ is the normalization constant dependent on $N$ and
$n$, see Eq.\eqref{eq:c0} and
\begin{align}\label{eq:duz}
d\mu(Z,Z^\dagger)=\frac{dZdZ^\dagger}{\det^{\frac{N}{2}+n-1}(1+ZZ^\dagger)}\;.
\end{align}
Integral on the left-hand side of Eq.\eqref{eq:fcft} is over the
real orthogonal group $\mathrm{O}(N)$. Here, $\psi^i_a$ and
$\bar{\psi}^i_a$, $i=1,\dots,N$ and $a=1,\dots,n$ are Grassmann
variables. Note that the $\bar{\psi}$'s are not necessarily related
to $\psi$'s by complex conjugation. That is, one can replace
$\bar{\psi}^i_a$ with an arbitrary set of independent Grassmann
variables which are not related to $\psi$ by any operation and the
identity will still hold. The integral on right-hand side of
Eq.\eqref{eq:fcft} is over complex skew-symmetric matrices of
dimension $n\times n$.

\bigskip

\noindent{{Bosonic O$(N)$ CFT}}
\begin{align}\label{eq:bcft}
\int_{\mathrm{O}(N)} dO\exp(\bar{\phi}^i_a O_{ij} \phi^j_a)=C_0^B\int_{1-ZZ^\dagger>0}
d\mu(Z,Z^\dagger)
\exp\frac{1}{2}\bigg(\bar{\phi}^i_aZ_{ab}\bar{\phi}^i_b+\phi^i_aZ_{ab}^\dagger\phi^i_b\bigg)\;,
\end{align}
where
\begin{align}
d\mu(Z,Z^\dagger)={\det}^{\frac{N}{2}-n-1}(1-ZZ^\dagger)dZdZ^\dagger\;.
\end{align}
Integral on the left-hand side of Eq.\eqref{eq:bcft} is again over
the real orthogonal group $\mathrm{O}(N)$. Here, $\phi^i_a$ and
$\bar{\phi}^i_a$, $i=1,\dots,N$ and $a=1,\dots,n$ are complex
variables. The normalization constant $c_0^B$ is defined in
Eq.\eqref{eq:cb0}. Similar to the fermionic case, one can also
replace $\bar{\phi}^i_a$ with an arbitrary set of independent
complex variables and the identity will still hold. The integral on
the right-hand side of Eq.\eqref{eq:bcft} is over complex symmetric
matrices of dimension $n\times n$ such that $1-ZZ^\dagger$ is
positive definite.

\bigskip

\noindent{{Supersymmetric O$(N)$ CFT}}
\begin{align}\label{eq:scft}
\int_{\mathrm{O}(N)} dO\exp(\bar{\psi}^i_a O_{ij} \psi^j_a)=
\int_{M_B\times M_F} d\mu(Z,\tilde{Z})
\exp\frac{1}{2}\bigg(\bar{\psi}^i_aZ_{ab}\bar{\psi}^i_b+\psi^i_a(-1)^{|a|}\tilde{Z}_{ab}\psi^i_b\bigg)\;.
\end{align}
As before, the integration on the left-hand side is over
$\mathrm{O}(N)$ and  $i,j=1,\dots,N$. Here $\psi$ and $\bar{\psi}$
are graded vectors whose elements $\psi^i_a$ and $\bar{\psi}^i_a$
are bosonic when $a=(\alpha, B)$ and fermionic when $a=(\alpha, F)$,
where $\alpha=1,\dots, n$. On the right-hand side, $Z$ and
$\tilde{Z}$ are $2n\times2n$ dimensional supermatrices subject to
the following condition
\begin{align}
Z=Z^T\sigma, \ \ \ \tilde{Z}=\sigma\tilde{Z}^T\;,
\end{align}
where $\sigma$ is the superparity defined as
$\sigma=\left(\begin{array}{cc} I_n & 0\\ 0 &
-I_n\end{array}\right)$. Here, $Z$ and $\tilde{Z}$ parameterize the
Riemannian symmetric superspace CI$|$DIII defined as
$\mathrm{OSp}(2n|2n)/\mathrm{GL}(n|n)$, such that
\begin{align}
M_B=\mathrm{Sp}(2n,,\mathbb{R})/\mathrm{U}(n), \ \ \ M_F=\mathrm{SO}(2n)/\mathrm{U}(n)\;.
\end{align}
We defined $d\mu(Z,\tilde{Z})=dZd\tilde{Z}\
\sdet^{\frac{N}{2}}(1-\tilde{Z}Z)$, where $dZd\tilde{Z}$ is the flat
Berezin measure on the space of supermatrices $Z$ and $\tilde{Z}$.
As usual, SDet denotes the superdeterminant. The integration domain
$M_B\times M_F$ is fixed by
\begin{align}
Z_{BB}=\tilde{Z}_{BB}^\dagger, \ \ \ Z_{FF}=-\tilde{Z}_{FF}^\dagger\;.
\end{align}
The symbol $|a|$ denotes the
parity of index $a$, $|a|=0$ when $a$ is bosonic and $|a|=1$ when $a$ is fermionic.

With the fermionic CFT, we calculated the ensemble average of
spectral determinants of certain real random matrices. The first one
is the positive integer moments of modulus of the characteristic
polynomial of matrices of the form $G\mathrm{O}(N)$, where $G$ is
diagonal. Its average over the orthogonal group O$(N)$ can be
written as an integral of products of  Pfaffians,
\begin{align}\label{eq:go}
\langle|z-GO|^{2m}\rangle_{\mathrm{O}(N)}=& \int_{\mathrm{O}(N)}dO\ {\det}^m\left[(z-GO)(z-GO)^\dagger\right]\nn\\
=&\ \mathrm{const.} \int_{Z=-Z^{\mathrm{T}}} d\mu(Z,Z^\dagger)
\prod_{i=1}^N {\mathrm{pf}}\!\left[\begin{array}{cc}g_i^2Z & \mathscr{Z}\otimes I_m\\
- \mathscr{Z}\otimes I_m  & Z^\dagger\end{array}\right]\;.
\end{align}
Here we introduced $2\times2$ matrix $\mathscr{Z}=\diag(z,\bar{z})$
and the integration measure is defined in Eq.\eqref{eq:duz}. In the
second example, we calculated the ensemble average of
characteristic polynomials of real random matrices $A$ from the
Jacobi ensemble,
\begin{align}
I_W(\lambda,\gamma)=\int dAdA^T {\det}^a(AA^T){\det}^b(1-AA^T)\det(\lambda-A)\det(\gamma-A^T)\;,
\end{align}
where $a$ and $b$ are non-negative integers. This average is again written in terms of Pfaffians,
\begin{align}\label{eq:even}
I_W(\lambda,\gamma)={\mathrm{const.}}\int_0^1 dr \frac{1}{(1+r)^{N+2}}\
\mathrm{pf}\;[\alpha_{i,j}]_{i,j=0,\dots,2s-1}\;,\ \ \ {\mathrm{when}}\ \ N=2s\;.
\end{align}
and
\begin{align}\label{eq:odd}
I_W(\lambda,\gamma)={\mathrm{const.}}\int_0^1 dr \frac{1}{(1+r)^{N+2}}\
\mathrm{pf}\!\left[\begin{array}{cc}\alpha_{i,j} & k_i(a,b;1)\\
-k_j(a,b;1)& 0\end{array}\right]_{i,j=0,\dots,2s}\hspace*{-12mm},\ \ {\mathrm{when}} \ N=2s+1\;.
\end{align}
Here $\alpha_{ij}$ is defined in Eq.\eqref{eq:alpha} and $k_i$ is defined in Eq.\eqref{eq:k}.

This paper is organized as follows. In section 2 we give a detailed
proof of the fermionic CFT and then outline the derivations of the
bosonic and supersymmetric CFT. In section 3, we calculate the
ensemble average of characteristic polynomials as applications of
the fermionic CFT. A summary is given in section 4.

\section{Proof of colour-flavor transformation}
We derive the $\mathrm{O}(N)$ colour-flavor transformations in this section.
A detailed derivation is given for the fermionic case. Derivations for the bosonic
and supersymmetric cases are given in more brief way with only important issues addressed,
whereas details are referred to either the fermionic case or literatures.

In this paper, all the three types of colour-flavor transformations are established by Howe's
dual pair theory, see \cite{cft, talk} and references therein. It is worth mentioning that in certain
cases symmetric polynomials can also be used to derive these transformations or even transformations
with different forms \cite{my,yb06}. However, it turns out that here the derivations based on
dual pair theory are more convenient.

\subsection{Fermionic $\mathrm{O}(N)$ CFT}
To establish the fermionic colour-flavor transformation over
$\mathrm{O}(N)$, we first derive the transformation over the special
orthogonal group $\mathrm{SO}(N)$,
\begin{align}\label{eq:soncft}
\int_{\mathrm{SO(N)}}\!\!\!\!\! d\mu(O)\exp(\bar{\psi}^i_a O_{ij} \psi^j_a)=C_0\!\int_{Z=-Z^{\mathrm{T}}}\!\!\!
\!\!\! d\mu(Z,Z^\dagger)
\exp\frac{1}{2}\bigg(\bar{\psi}^i_aZ_{ab}\bar{\psi}^i_b+\psi^i_aZ_{ab}^\dagger\psi^i_b\bigg)
(1+K\det\mathcal{M})\;,
\end{align}
where $C_0^F$ and $K\!=\!\frac{C_1^F}{C_0^F\lambda^n_1 N!}$ are constants defined later and $\mathcal{M}$
is an $N\times N$ matrix defined as
$\mathcal{M}_{ij}=\left[\bar{\psi}^i_a(1+ZZ^\dagger)_{ab}\psi^j_b\right]$.\\

\noindent{\bf{Remark}}: To get Eq.\eqref{eq:fcft}, we exploit the fact that
$\mathrm{O}(N)\!=\!\mathrm{O}_+(N)\!\oplus\!\mathrm{O}_-(N)$, where $\mathrm{O}_+\!\cong\! \mathrm{SO}(N)$
and $\mathrm{O}_-$ is a rotation followed by a 'reflection' $R$, which
can be chosen as $R=\diag(I_{N-1},-1)$. Note that $\bar{\psi}^iR_{ij}$ flops the sign of
all $\bar{\psi}^N$'s therefore inverts the sign of $\det\mathcal{M}$. Hence, the parts
containing $\det\mathcal{M}$ are cancelled when we combine the contributions from the normal
rotation $\mathrm{SO}(N)$ and the improper rotation $R\cdot\mathrm{SO}(N)$.
Which proofs Eq.\eqref{eq:fcft}. In fact, it turns out that one can 'naively' use a
similar mapping as defined above Eq.\eqref{eq:po} for $\mathrm{O}(N)$ and get Eq.\eqref{eq:fcft}
with less effort. However, due to the fact that $\det O=\pm1$ for $O\in\mathrm{O}(N)$
this kind of mapping is mathematically better defined for $\mathrm{SO}(N)$.

In the following paragraphs, we will mainly follow the method of
paper \cite{bt}. Introduce fermionic creation and annihilation
operator $\bar{f}^i_a$ and $f^i_a$, where $i=1,\dots,N$ and
$a=1,\dots,n$. As usual, we borrow the terminology from lattice
gauge theory where the upper indices are referred to as 'colour' and
the lower indices are referred to as 'flavor'. This set of operators
satisfy the canonical fermion anticommutation relations
$\{f^i_a,\bar{f}^j_b\}=\delta^{ij}\delta_{ab}$ and
$\{f^i_a,f^j_b\}=\{\bar{f}^i_a,\bar{f}^j_b\}=0$. These operators
construct a Fock space $\mathscr{F}_F$ for a fermionic system. Let
$|0\rangle$ be the vacuum state, then we have $f^i_a|0\rangle=0$.
The quadratic operators $\bar{f}^i_a\bar{f}^j_b$, $f^i_af^j_b$ and
$\bar{f}^i_af^j_b-f^i_a\bar{f}^j_b$ therefore define a
representation of the Lie algebra $\mathrm{so}(2nN,\mathbb{C})$.
This algebra has to two commuting subalgebras. Which are
$\mathrm{so}(2n,\mathbb{C})$ generated by $\bar{f}^i_a\bar{f}^i_b$,
$f^i_af^i_b$ and $\bar{f}^i_af^i_b-f^i_a\bar{f}^i_b$ and
$\mathrm{so}(N,\mathbb{C})$ generated by $\bar{f}^i_af^j_a+f^i_a
\bar{f}^j_a$.

To derive colour-flavor transformation we need to construct two
types of projection operators $\hat{P}_C$ and $\hat{P}_F$ which
project states to a subspace of $\mathscr{F}_F$, named colour-single
space, which is invariant under the $\mathrm{SO}(N)$ group action
$O\mapsto T_O:=\exp(\bar{f}^i_a(\ln O)_{ij}f^j_a)$. As usual, we
define
\begin{align}\label{eq:po}
\hat{P}_C=\int_{\mathrm{SO}(N)}dO\ T_O\;.
\end{align}
Note that for $\mathrm{SO}(N)$, the colour-single space has two
disconnected components, each of which carries an irreducible
representation of $\mathrm{SO}(N)$. One of them contains the vacuum
state $|\psi_0\rangle=|0\rangle$ and is in addition $\mathrm{O}(N)$
invariant, and the other one contains the state
$|\psi_1\rangle:=\bar{f}^1_1\bar{f}^2_1\dots\bar{f}^N_1|0\rangle$.
Acting on the vacuum and $|\psi_1\rangle$ by the operators
$\bar{f}^i\bar{f}^i$ span each colour-single subspace. Therefore, we
have $\hat{P}_F=\hat{P}_{F_0}+\hat{P}_{F_1}$.

It is convenient to define the notation
\begin{align}
\bar{c}^i_A=\left\{\begin{array}{ll}\bar{f}^i_A & A=1,\dots,n \\
f^i_{A-n} & A=n+1,\dots,2n\end{array}\right.
\ \ \ \mathrm{and}\ \ \
c^i_A=\left\{\begin{array}{ll}f^i_A & A=1,\dots,n \\
\bar{f}^i_{A-n} & A=n+1,\dots,2n\end{array}\right.\;.
\end{align}
For $g\!\in\!\mathrm{SO}(2n)$, It is direct to check the mapping
$g\mapsto T_g:=\exp(\frac{1}{2}\bar{c}^i_A(\ln g)_{AB}c^i_B)$
constructs a representation of $\mathrm{SO}(2n)$, such that
$T_g\bar{c}^i_AT_{g}^{-1}=\bar{c}^i_Bg_{BA}$. An element of
$\mathrm{SO}(2n)$ can be parameterized as \cite{pere}
\begin{align}
g=\left(\begin{array}{cc}U&V \\ \bar{V}&\bar{U}\end{array}\right), \ \
{\mathrm{where}}\ \ UU^\dagger+VV^\dagger=I,\ \ {\mathrm{and}}\ \ UV^T+VU^T=0\;.
\end{align}
Note that this $G$ is a subgroup of $\mathrm{U}(2n)$ that is
isomorphic to $\mathrm{SO}(2n)$. It is clear that $G$ has a
$\mathrm{U}(n)$ subgroup $G\supset
H=\diag(U,\bar{U})\cong\mathrm{U}(n)$. We parametrize the coset
space $G/H$ as
\begin{align}
G/H=\left(\begin{array}{cc}(1+ZZ^\dagger)^{-\frac{1}{2}} & Z(1+Z^\dagger Z)^{-\frac{1}{2}}\\
Z^\dagger(1+ZZ^\dagger)^{-\frac{1}{2}} & (1+Z^\dagger Z)^{-\frac{1}{2}}
\end{array}\right)\;,
\end{align}
where $Z=U^{-1}V$ are complex skew-symmetric matrices of dimension $n\times n$.
\begin{align}
\hat{P}_{F}=\sum_{\sigma=0,1}\hat{P}_{F_\sigma}=\sum_{\sigma=0,1}C_\sigma\int_{\mathrm{SO}(2n)}\!\!
dg\ T_g|\psi_\sigma\rangle\langle \psi_\sigma|T_g^{-1}\;,
\end{align}
where $C^F_\sigma$ is defined by the normalization condition
$\langle\psi_\sigma|\hat{P}_{F_\sigma}|\psi_\sigma\rangle=1$. Define
$|Z\rangle=\exp(\frac{1}{2}\bar{f}^i_aZ_{ab}\bar{f}^i_b)|0\rangle$
then it is direct to check
\begin{align}
\hat{P}_{F_0}=C_0^F\int_{\mathrm{SO}(2n)}\!\!
dg\ T_g|\psi_0\rangle\langle \psi_0|T_g^{-1}=C_0\int_{Z=-Z^T}\!\! d\mu(Z,Z^\dagger)\
|Z\rangle\langle Z|\;.
\end{align}
Using the method in chapter 2 of \cite{hua}, we get the
normalization factor
\begin{align}\label{eq:c0}
C_0^F=\pi^{-\frac{n(n-1)}{2}}\prod\limits_{i=1}^{n-1}\frac{\Gamma(N+2i)}{\Gamma(N+i)}\;.
\end{align}
Note that we choose the group volume of $\mathrm{SO}(N)$ to be 1. Follow the method
in \cite{bt}, ${C^F_1}^{-1}$ equals to the dimension of the representation of $\mathrm{SO}(2n)$
defined by the Young diagram with 1 row and N columns.

Now define $|\psi\rangle=\exp(\bar{f}^i_a\psi^i_a)|0\rangle$, then the colour-flavor transformation
follows from the identity $\langle\psi|\hat{P}_C|\psi\rangle=\langle\psi|\hat{P}_F|\psi\rangle$.
It is straightforward to show
\begin{align}
\langle\psi|\hat{P}_C|\psi\rangle=\int_{\mathrm{SO(N)}} dO\exp(\bar{\psi}^i_a O_{ij} \psi^j_a)\;,
\end{align}
and
\begin{align}
\langle\psi|\hat{P}_{F_0}|\psi\rangle=C_0^F\int_{Z=-Z^{\mathrm{T}}} d\mu(Z,Z^\dagger)
\exp\frac{1}{2}\bigg(\bar{\psi}^i_aZ_{ab}\bar{\psi}^i_b+\psi^i_aZ_{ab}^\dagger\psi^i_b\bigg)\;.
\end{align}
Follow the method in \cite{bt} again, we get
\begin{align}
\langle\psi|\hat{P}_{F_1}|\psi\rangle
=\frac{C_1^F}{\lambda^n_1 N!}\int_{Z=-Z^{\mathrm{T}}} d\mu(Z,Z^\dagger)
\exp\frac{1}{2}\bigg(\bar{\psi}^i_aZ_{ab}\bar{\psi}^i_b+\psi^i_aZ_{ab}^\dagger\psi^i_b\bigg)
\det\left[\bar{\psi}^i_a(1+ZZ^\dagger)_{ab}\psi^j_b\right]
\end{align}
where $\lambda^n_1$ is the dimension of the representation of $\mathrm{U}(n)$ defined by
the Young diagram with 1 row and N columns. Combining the above formula, we complete the
proof of Eq.\eqref{eq:soncft}

\subsection{Bosonic $\mathrm{O}(N)$ CFT}
The bosonic colour-flavor transformation is derived with almost the
same method \cite{zirn} but with the following few changes. We
construct a bosonic Fock space $\mathscr{F}_B$ with bosonic
operators. The two commuting algebras are
$\mathrm{so}(N,\mathbb{C})$ and $\mathrm{sp}(2n,\mathbb{C})$
\cite{zirn}. Define
\begin{align}
\bar{c}^i_A=\left\{\begin{array}{ll}\bar{b}^i_A & A=1,\dots,n \\
b^i_{A-n} & A=n+1,\dots,2n\end{array}\right.
\ \ \ \mathrm{and}\ \ \
c^i_A=\left\{\begin{array}{ll}b^i_A & A=1,\dots,n \\
-\bar{b}^i_{A-n} & A=n+1,\dots,2n\end{array}\right.\;.
\end{align}
For $g\!\in\!\mathrm{Sp}(2n,\mathbb{R})$, it is straightforward to
check that the mapping $g\mapsto
T_g:=\exp(\frac{1}{2}\bar{c}^i_A(\ln g)_{AB}c^i_B)$ constructs a
representation of $\mathrm{Sp}(2n,\mathbb{R})$, such that
$T_g\bar{c}^i_AT_{g}^{-1}=\bar{c}^i_Bg_{BA}$. An element of
$\mathrm{Sp}(2n,\mathbb{R})$ can be parameterized as
\begin{align}
g=\left(\begin{array}{cc}U&V \\ \bar{V}&\bar{U}\end{array}\right), \ \
{\mathrm{where}}\ \ UU^\dagger-V^T\bar{V}=I,\ \ {\mathrm{and}}\ \ U^\dagger V=V^T\bar{U}\;.
\end{align}
Here $G$ is a subgroup of $\mathrm{U}(n,n)$ that is isomorphic to
$\mathrm{Sp}(2n,\mathbb{R})$. Note that $G$ has a $\mathrm{U}(n)$
subgroup. The coherent states are similarly defined as before,
$|Z\rangle=\exp(\frac{1}{2}\bar{b}^i_aZ_{ab}\bar{b}^i_b)|0\rangle$
and $|\phi\rangle=\exp(\bar{b}^i_a\phi^i_a)|0\rangle$. Following the
same procedure as before and \cite{zirn}, we get Eq.\eqref{eq:bcft}.

In order to fix the normalization factor, choosing the group volume
of $\mathrm{O}(N)$ to be 1 and using Eq. of \cite{hua}, we get
\begin{align}\label{eq:cb0}
C^B_0=\pi^{-\frac{n(n+1)}{2}}\frac{N-2n}{2}\prod\limits_{i=1}^{n-1}
\frac{(\frac{N}{2}-i)\Gamma(N-1-i)}{\Gamma(N-1-2i)}\;.
\end{align}

\subsection{Supersymmetric $\mathrm{O}(N)$ CFT}
When $\psi$ and $\bar{\psi}$ are graded vectors, we need to extend the flavor group
to incorporate this 'supersymmetry'. The Lie superalgebra we need satisfies the following
condition
\begin{align}
Q=-\gamma Q^T\gamma\;.
\end{align}
Let $\sigma$'s being the Pauli matrices, $\gamma$ can be defined as
\begin{align}
\gamma=\left(\begin{array}{cc} \mathrm{i}\sigma_y & 0\\ 0 & \sigma_x\end{array}\right)\otimes I_n\;.
\end{align}
Therefore, $Q$ defines the Lie superalgebra $\mathrm{osp}(2n|2n)$ whose boson-boson block
is the Lie algebra $\mathrm{sp}(2n)$ and fermi-fermi block is $\mathrm{so}(2n)$. The
remaining derivation closely follows \cite{cft}. To do a simple check, setting the bosonic
(or fermionic) degree of freedom on both sides of Eq.\eqref{eq:scft} to zero and choose corresponding
integral measure, we can recover the fermionic (or bosonic) CFT Eq.\eqref{eq:fcft}.

\section{Applications in real random matrices}
In this section, we calculate the averaged characteristic polynomials of real random matrices
from certain ensembles. These calculations are simple applications of the fermionic O$(N)$ CFT.

\subsection{Modulus square of characteristic polynomials of certain type of real random
matrices averaged over O$(N)$}
Let $z$ be a complex number and $G$ be a real diagonal matrix. We calculate the following
quantity,
\begin{align}
F_{G}(z)=\langle|z-GO|^{2m}\rangle_{\mathrm{O}(N)}=\int_{\mathrm{O}(N)}dO\
{\det}^m\left[(z-GO)(z-GO)^\dagger\right]\;.
\end{align}
Note that $\langle|z-GU|^{2m}\rangle_{\mathrm{U}(N)}$, where the
average is over unitary group, has been calculated with the method
of symmetric functions in \cite{yb05}. And it is not hard to show
that the method we will use in this section can be applied to the
unitary case as well, in which case one will use the U$(N)$
fermionic CFT.

Introducing Grassmann vectors, $\eta^i_a$ and $\xi^i_a$, $i=1,\dots,N$ and $a=1,\dots,m$,
we can re-write Eq.\eqref{eq:gauss} as
\begin{align}\label{eq:start}
F_{G}(z)=& \int_{\mathrm{O}(N)} dO \int d\bar{\eta}d\eta d\bar{\xi}d\xi\
\mathrm{e}^{-z\bar{\eta}^i_a\eta^i_a-\bar{z}\bar{\xi}^i_a\xi^i_a
+\bar{\eta}^i_ag_iO_{ij}\eta^j_a+\xi^i_aO_{ij}^Tg_j\bar{\xi}^j_a}\nn\\
=&  \int d\bar{\eta}d\eta d\bar{\xi}d\xi\  \mathrm{e}^{-z\bar{\eta}^i_a\eta^i_a-\bar{z}\bar{\xi}^i_a\xi^i_a}
\int_{\mathrm{O}(N)} dO \exp(\bar{\psi}^i_aO_{ij}\psi^j_a)\;,
\end{align}
where we introduced the composite notation $(\bar{\psi}^i_{1a},\bar{\psi}^i_{2a})=(\bar{\eta}^i_ag_i,-\xi^i_ag_i)$
(no summation) and $({\psi}^i_{1a},{\psi}^i_{2a})=(\eta^i_a, \bar{\xi}^i_a)$ and changed the order of
integration. Defining the $2m\times2m$ complex skew-symmetric matrix $Z$ and applying the fermionic
CFT Eq.\eqref{eq:fcft}, we get
\begin{align}\label{eq:end}
F_{G}(z)=&\ C_0^F \int d\bar{\eta}d\eta d\bar{\xi}d\xi\ \mathrm{e}^{-z\bar{\eta}^i\eta^i-\bar{z}\bar{\xi}^i\xi^i}
\int_{Z=-Z^{\mathrm{T}}} d\mu(Z,Z^\dagger)
\exp\frac{1}{2}\bigg(\bar{\psi}^i_aZ_{ab}\bar{\psi}^i_b+\psi^i_aZ_{ab}^\dagger\psi^i_b\bigg)\;.
\end{align}
Next, we change the order of integration and use the standard Gaussian
integral formula for Grassmann vectors. The final result is written in
terms of Pfaffians and given in Eq.\eqref{eq:go}.

When $m=1$, we can integrate over $Z$ explicitly since
\begin{align}
Z=\left(\begin{array}{cc}  0 & a\\ -a & 0\end{array}\right)\;.
\end{align}
By Eq.\eqref{eq:end}, we get
\begin{align}
F_{G}(z)=&\ \langle|z-GO|^2\rangle_{\mathrm{O}(N)}
=\ \mathrm{const.} \int dad\bar{a}\frac{1}{(1+a\bar{a})^{N+2}}
\prod_{i=1}^N(|z|^2+a\bar{a}g_i^2)\nn\\
=&\ \mathrm{const.} \sum_{l=0}^N \frac{1}{C_N^l}|z|^{2(N-l)}S^l(G^2)\;,
\end{align}
Here, $C_N^l=N!/l!(N-l)!$ is the binomial function and $S^l(G^2)$ denotes the l-th order elementary
symmetric polynomials of $g_i^2$'s, i.e. $S^0(G^2)=1$, $S^1(G^2)=\sum_i g_i^2$,
$S^2(G^2)=\sum_{i<j}g_i^2g_j^2$, etc.

\subsection{Characteristic polynomials averaged over Jacobi ensemble}
As another simple application of the transformations we derived in the previous section,
we calculate the following average,
\begin{align}\label{eq:gauss}
\int dAdA^T W(AA^T) \det(\lambda-A)\det(\gamma-A^T)\;.
\end{align}
Where $W(AA^T)$ is an arbitrary invariant ensemble which depends on singular values of $A$ only,
i.e. $W(A)=W(OAO')=W(G)$, where $O$ and $O'$ are arbitrary elements in $\mathrm{O}(N)$ and $G$
are the singular values of $A$ defined by $A=O_1GO_2$. In particular, we are interested in
ensembles whose potentials are also separable functions of $G$, i.e.
\begin{align}\label{eq:w}
W(AA^T)=\prod_i W(g_i^2)\;.
\end{align}

First, we make singular decomposition $A=O_1GO_2$, where $G=\diag(g_1,\dots,g_N)\ge0$.
\begin{align}
\int dAdA^T W(AA^T)=\int_{\mathrm{O}(N)} dO_1dO_2 \int \prod_i W(g_i^2)dg_i\prod_{i<j}|g_i^2-g_j^2|\;.
\end{align}
By invariance of Haar measure, we have $\det(\lambda-O_1GO_2)=\det(\lambda-O_2O_1G)=\det(\lambda-OG)$.
Introducing Grassmann vectors as before, we can re-write Eq.\eqref{eq:gauss} as
\begin{align}
\int_{\mathrm{O}(N)} dO \int \prod_i W(g_i^2)\ dg_i\prod_{i<j}|g_i^2-g_j^2|
\int d\bar{\eta}d\eta d\bar{\xi}d\xi\
\mathrm{e}^{-\lambda\bar{\eta}^i\eta^i-\gamma\bar{\xi}^i\xi^i+\bar{\eta}^iO_{ij}g_j\eta^j-\xi^iO_{ij}g_j\bar{\xi}^j}\;.
\end{align}
Use the same method as in Eq.\eqref{eq:start}-Eq.\eqref{eq:end}, we get for separable
invariant potentials Eq.\eqref{eq:w},
\begin{align}\label{eq:main}
I_W(\lambda,\gamma)=&\int dAdA^T W(AA^T)\det(\lambda-A)\det(\gamma-A^T)\nn\\
=&\mathrm{const.}\int_0^1 dr \frac{1}{(1+r)^{N+2}} \int
\prod_{i<j}|g_i^2-g_j^2|\ \prod_i \left(\lambda\gamma+rg_i^2\right)W(g_i^2) dg_i\;.
\end{align}
One benefit of using singular values instead of eigenvalues is that all integral
variables are real non-negative. From the above formula it is clear that when
$\gamma=\bar{\lambda}$ the integral on the left-hand side depends only on $|\lambda|^2$.

Note that Eq.\eqref{eq:main} applies to all ensembles of real random matrices
satisfying Eq.\eqref{eq:w}. These ensembles include $W(AA^T)=\mathrm{e}^{-\tr V(AA^T)}$, which becomes the
Ginibre ensemble when $V(x)=\frac{1}{2}x$, which has been studied intensively.
In the remaining part of this section, we consider the Jacobi ensemble,
\begin{align}
W(x)=x^a(1-x)^b\;.
\end{align}

To proceed, we use the method in Chapter 5 of \cite{mehta}.
\begin{align}\label{eq:meh}
\int_0^1 \prod_{i<j}|g_i^2-g_j^2|\
 \prod_i\left(1+rg_i^2\right)W(g_i^2) dg_i
=N!\!\!\!\!\!\int\limits_{0<g_N\dots<g_1<1}\!\!\!\!\! \prod_i dg_i
\det\left[W(g_i^2)R_{j-1}(g_i^2)(1+rg_i^2)\right]\;.
\end{align}
where $R_{j}(x)$ can be arbitrary monic $j$ order polynomials of x. Here we choose $R_j(x)=x^j$.
Define the following functions
\begin{align}
h(a,b;x)=\int_0^x dg\ g^{2a}(1-g^2)^b=\frac{1}{2}\Gamma(b+1)\Gamma(a+\frac{1}{2})\sum_{i=0}^b
\frac{x^{2(a+i)+1}(1-x^2)^{b-i}}{\Gamma(b-i+1)\Gamma(a+i+\frac{3}{2})}\;,
\end{align}
and
\begin{align}\label{eq:k}
k_i(a,b;x)=\int_0^x dg\ (1+rg^2)g^{2(a+i)}(1-g^2)^b=h(a+i,b;x)+\frac{r}{\lambda\gamma}h(a+i+1,b;x)\;,
\end{align}
Let $B(x,y)=\frac{\Gamma(x)\Gamma(y)}{\Gamma(x+y)}$ be the Beta function. When $N=2s$, introduce the
skew-symmetric matrix $\alpha$,
\begin{align}\label{eq:alpha}
\alpha_{ij}=&\int_0^1 dg\ (1+rg^2)g^{2a}(1-g^2)^b\left(g^{2i}k_j(g)-g^{2j}k_i(g)\right)\nn\\
=& \frac{1}{4}\bigg\{ \frac{B(b+1,a+j+\frac{1}{2})}{a+b+j+\frac{3}{2}}\sum_{l=0}^b
\frac{B(2b-l+1,2a+1+i+j+l)}{B(b-l+1,a+j+l+\frac{3}{2})}+\nn\\
&\ \ \ \ \frac{r}{\lambda\gamma} \frac{B(b+1,a+j+\frac{1}{2})}{a+b+j+\frac{3}{2}}\sum_{l=0}^b
\frac{B(2b-l+1,2a+2+i+j+l)}{B(b-l+1,a+j+l+\frac{3}{2})}+\nn\\
&\ \ \ \  \frac{r}{\lambda\gamma}\frac{B(b+1,a+j+\frac{3}{2})}{a+b+j+\frac{5}{2}}\sum_{l=0}^b
\frac{B(2b-l+1,2a+2+i+j+l)}{B(b-l+1,a+j+l+\frac{5}{2})}+\nn\\
&\ \ \ \ \left(\frac{r}{\lambda\gamma}\right)^2 \frac{B(b+1,a+j+\frac{3}{2})}{a+b+j+\frac{5}{2}}\sum_{l=0}^b
\frac{B(2b-l+1,2a+3+i+j+l)}{B(b-l+1,a+j+l+\frac{5}{2})}-i\leftrightarrow j\bigg\}\;.
\end{align}
By Eq.(5.5.8) and Eq.(5.5.9) of \cite{mehta}, we get Eq.\eqref{eq:even} and Eq.\eqref{eq:odd}.

Similarly, for $W(x)=\exp(-\frac{1}{2}x)$, we can check that Eq.\eqref{eq:main}
and Eq.\eqref{eq:meh} gives the same results in \cite{som}
\begin{align}
\int dAdA^T \mathrm{e}^{-\frac{1}{2}\tr AA^T}\det(\lambda-A)\det(\gamma-A^T)
={\mathrm{const.}}\sum_{n=0}^N\frac{(\lambda\gamma)^n}{n!}\;.
\end{align}

\section{Conclusions}
In this paper, we derived the O$(N)$ colour-flavor transformations.
We believe that these transformations will be useful in the study of
real random matrices, just as the U$(N)$ CFT are for complex random
matrices. As  a simple application, we calculated averaged
characteristic polynomials for two types of ensembles,  which, to
the best of our knowledge, have not been calculated before.

We have only showed examples with the fermionic CFT whereas applications of the bosonic and
supersymmetric CFTs will be postponed to future works. In which, as usual, we need to pay
attention to analytical issues, especially when we need to change the order of integrations,
\cite{yb05,me}. Also, by its nature the supersymmetric CFT often requires more work.
However, when $n=1$ Eq.\eqref{eq:scft} enjoys the simplicity that
\begin{align}
Z=\left(\begin{array}{cc}  a & \sigma \\ -\sigma & 0\end{array}\right)\ \ \ {\mathrm{and}}
\ \ \
\tilde{Z}=\left(\begin{array}{cc}  \bar{a} & \rho \\ \rho & 0\end{array}\right)\;.
\end{align}
This is because $\mathrm{O}(2)/\mathrm{U}(1)$ consists of only single points.

\section*{Acknowledgements}
This research was completed during the 2008 programme "Anderson
Localization: 50 years after" at the Issac Newton Institute for
Mathematical Sciences where both authors were supported by visiting
Fellowships. We gratefully acknowledge Prof. M.R. Zirnbauer for
clarifying discussions.

\end{document}